\documentclass[12pt,preprint]{aastex}
\usepackage{epsfig}
\received{}
\accepted{}
\journalid{}{}
\articleid{}{}

\shorttitle{Optical ground versus space comparison}
\shortauthors{Mountain et al.}

\newcommand{\etal}{{et al.~}}
\newcommand{\lta}{\lesssim}
\newcommand{\gta}{\gtrsim}
\newcommand{\onem}{$\sim 1 \mu$m}
\newcommand{\pap}{\noindent\hangindent=0.3truecm\hangafter=1}
\newcommand{\papp}{\indent\hspace{8.0truecm}\hangindent=9.1truecm\hangafter=1}

\begin{document}

\title{Comparison of optical observational capabilities\\ 
       for the coming decades: ground versus space}

\author{Matt Mountain\altaffilmark{1,2},\\
        Roeland van der Marel\altaffilmark{1},
        Remi Soummer\altaffilmark{1},\\ 
        Anton Koekemoer\altaffilmark{1},
        Harry Ferguson\altaffilmark{1},
        Marc Postman\altaffilmark{1},\\
        Donald T.~Gavel\altaffilmark{6},
        Olivier Guyon\altaffilmark{4},
        Douglas Simons\altaffilmark{3},
        Wesley A.~Traub\altaffilmark{5}}

\altaffiltext{1}{Space Telescope Science Institute, 3700 San Martin Drive, 
                 Baltimore, MD 21218
                 \vspace{-0.2truecm}}

\altaffiltext{2}{Visiting Professor, Dept.~Physics, Univ.~of Oxford,
                 1 Keble Road, Oxford OX1 3NP, United Kingdom
                 \vspace{-0.2truecm}}

\altaffiltext{3}{Gemini Observatory, 670 N. A'ohoku Pl., Hilo, HI 96720
                 \vspace{-0.2truecm}}

\altaffiltext{4}{Subaru Telescope, 650 N. A'ohoku Pl., Hilo, HI 96720
                 \vspace{-0.2truecm}}

\altaffiltext{5}{JPL, M/S 301-451, 4800 Oak Grove Drive, Pasadena, CA 91109
                 \vspace{-0.2truecm}}

\altaffiltext{6}{UCO/Lick Observatory, UCSC, 1156 High Street, CfAO
                 Building Santa Cruz, CA 95064
                 \vspace{-0.2truecm}}

\baselineskip=12pt


\begin{abstract}
\null\vspace{-0.8truecm}\baselineskip=12pt Ground-based adaptive optics (AO)
in the infrared has made exceptional advances in approaching
space-like image quality at higher collecting area. Optical-wavelength
applications are now also growing in scope. We therefore provide here
a comparison of the pros and cons of observational capabilities from
the ground and from space at optical wavelengths. With an eye towards
the future, we focus on the comparison of a $\sim 30$m ground-based
telescope with an 8--16m space-based telescope. The parameters
relevant for such a comparison include collecting area, diffraction
limit, accessible wavelength range, background emission, atmospheric
absorption and extinction, Strehl ratio, field of view, temporal and
spatial PSF stability, and target accessibility in time and on the
sky. We review the current state-of-the-art in AO, and summarize the
expected future improvements in image quality, field of view,
contrast, and low-wavelength cut-off. We compare the depth that can be
reached for imaging and spectroscopy from the ground and from space in
the V and J bands. We discuss the exciting advances in extreme AO for
exoplanet studies and explore what the theoretical limitations in
achievable contrast might be. Our analysis shows that extreme AO
techniques face both fundamental and technological hurdles to reach
the contrast of $10^{-10}$ necessary to study an Earth-twin at 10
pc. Based on our assessment of the current state-of-the-art, the
future technology developments, and the inherent difficulty of
observing through a turbulent atmosphere, we conclude that there will
continue to be a strong complementarity between observations from the
ground and from space at optical wavelengths in the coming
decades. There will continue to be subjects that can {\it only} be
studied from space, including imaging and (medium-resolution)
spectroscopy at the deepest magnitudes, and the exceptional-contrast
observations needed to characterize terrestrial exoplanets and search
for biomarkers.\vfill\noindent additional contact information primary
author:\\ Matt Mountain, STScI, mmountain@stsci.edu, 410-338-4710
\end{abstract}

\addtocounter{page}{-1}


\section{Introduction}
\label{s:intro}

\vspace{-0.2truecm}

The future of optical observational astronomy looks bright, with
excellent prospects for the advent of larger aperture telescopes both
on the ground and in space. As the telescope size increases, so does
the cost and complexity of a project. The 2010 Decadal Survey will
therefore face difficult choices for ground- and space-based astronomy
investments in the coming decade. To facilitate these choices, it is
important to understand the relative advantages of ground- and
space-based facilities for observational astronomy. In particular, the
ever increasing capabilities of adaptive optics (AO) from the ground
motivate a critical side-by-side comparison of the observational
parameter space accessible to either type of facility.\looseness=-2

We focus here mostly on comparison of the capabilities of a $\sim 30$m
ground-based telescope with an 8--16m space-based telescope. Examples
of the former are the Thirty Meter Telescope (TMT; Stone \etal 2009)
and the Giant Magellan Telescope (GMT; McCarthy \etal 2009). Examples
of the latter are the Advanced-Technology Large-Aperture Space
Telescope (ATLAST; Postman \etal 2009) and the Terrestrial Planet
Finder Coronagraph (TPF-C; Levine \etal 2009). We compare only briefly
the capabilities of current ground-based 8--10m class telescopes to
smaller space-based telescopes such as the 2.4m Hubble Space Telescope
(HST), the 6.5m James Webb Space Telescope (JDEM), or the proposed
Joint Dark Energy Mission (JDEM/IDECS; Gehrels \etal 2009).

We restrict attention to the regime below {\onem}, for two
reasons. First, this is where future large space-based proposals are
likely to focus their attention; JWST will already provide an
important new large aperture facility for science longward of
{\onem}. Second, below {\onem} is where the future capabilities of AO
from the ground are most uncertain and least documented. AO
corrections in the infrared (IR) are typically more capable and less
technically challenging than in the optical (see
Section~\ref{s:AOtypes}). However, in comparing performance of space
versus ground in the IR, careful consideration must be given to the
role of background emission in achieving a given signal-to-noise ratio
$S/N$ (see Section~\ref{s:SN}).

We will not discuss here the relative cost for new optical observing
facilities on the ground or in space. Observational facilities in
space are generally more expensive to build and operate than those on
the ground. So they are usually pursued only if they open up parts of
parameter space that are not accessible from the ground. We highlight
which parts of observational parameter space are uniquely accessible
only from space. Scientific motivations for access to this parameter
space are briefly mentioned where relevant. Costs and science drivers
are discussed in detail in other submissions to the Decadal Survey.

\vspace{-0.9truecm}

\section{Characteristics of Observational Parameter Space}
\label{s:param}

\vspace{-0.4truecm}

\subsection{Telescope Size}
\label{ss:telsize}

\vspace{-0.3truecm}

One of the main advantages for ground-based facilities is that they
can generally be constructed with larger telescope diameters $D$ than
what is possible in space, which impacts two important characteristics:
the collecting area and the diffraction limit.

\vspace{-0.5truecm}

\paragraph{2.1.1. Collecting Area} The collecting area of a telescope scales 
as $D^2$. This implies that large-aperture ground-based telescopes are
generally able to collect more photons than space-based telescopes,
which is one of the factors that determines the achievable depth and
signal-to-noise ratio ($S/N$; see Section~\ref{s:SN}).

\vspace{-0.5truecm}

\paragraph{2.1.2. Diffraction Limit} The diffraction limit of a 
telescope scales as $\lambda/D$. This implies that large-aperture
ground-based telescopes generally have a smaller diffraction limit
than space-based telescopes. However, due to atmospheric turbulence it
is more difficult for ground-based telescopes to achieve image quality
near the diffraction limit than it is for space-based telescopes,
especially at optical wavelengths (see Sections~\ref{ss:imqual},
\ref{s:AOtypes} and~\ref{s:contrast}).

\vspace{-0.7truecm}

\subsection{Atmospheric Emission and Absorption}
\label{ss:atmos} 

\vspace{-0.2truecm}

One of the most fundamental advantages of space-based over
ground-based observations is the fact that light can be observed
unaffected by the atmosphere. This impacts several important
characteristics of the observational parameter space.

\vspace{-0.5truecm}

\paragraph{2.2.1. Accessible Wavelength Range and Absorption} Ultra-Violet 
observations are not possible at all from the ground, due to
atmospheric absorption. This blocks an entire wavelength range in
which many astrophysical problems can be uniquely studied. Even in the
optical there are spectroscopic absorption features (possibly
time-variable) associated with telluric bands. These need to be
corrected using observations of standard stars, which themselves are
only characterized to finite accuracy.

\vspace{-0.5truecm}

\paragraph{2.2.2. Background} The atmosphere creates a background in 
ground-based observations that needs to be subtracted. Figure~1
compares the background at Mauna Kea and L2. The shot noise from the
background affects the the achievable depth and $S/N$ (see
Section~\ref{s:SN}). In addition, background variations can lead to
systematic errors due to imperfect subtraction.\looseness=-2

\vspace{-0.5truecm}

\paragraph{2.2.3. Line Emission} In addition to a continuum background, 
ground-based observations accumulate photons from atmospheric emission
in well-defined bands with many narrow emission lines (see
Figure~1). The presence of these emission lines increases towards the
IR, and their strengths vary with time. This limits the ability to
study spectral features in astronomical objects at particular
wavelengths.

\vspace{-0.5truecm}

\paragraph{2.2.4. Atmospheric Extinction} Atmospheric extinction must 
be corrected on the basis of airmass estimates and standard star
observations, to obtain absolute photometry. The accuracy with which
these corrections can be done are critical for certain areas of
science.

\vspace{-0.8truecm}

\subsection{Image Quality}
\label{ss:imqual}

\vspace{-0.2truecm}

The image quality depends both on the diffraction limit of the
telescope (Section 2.1.2) and the ability of the telescope to reach
this diffraction limit.

\vspace{-0.5truecm}

\paragraph{2.3.1. Strehl Ratio (SR)} The SR is the ratio of the peak 
flux in the normalized point-spread function (PSF) to that for a
diffraction limited system. It (and other measures of image quality)
describe the extent to which the light is concentrated in the PSF core
as opposed to the wings. A space-based telescope is typically designed
to be diffraction limited (commonly defined as ${\rm SR} > 80$\%), at
some target wavelength driven by the science. By contrast, large
ground-based telescopes need an AO system that aims to optimize a
combination of the complimentary goals of high SR and large field of
view (see Sections~\ref{s:AOtypes} and~\ref{s:contrast}). Even when an
AO system achieves a diffraction-limed core, a low SR can severely
limit the science. For example, in a crowded field the PSF wings of
bright stars create an elevated background that drowns out the light
of fainter stars.

\vspace{-0.5truecm}

\paragraph{2.3.2. Field of View (FOV)} The FOV sizes for space- and 
ground-based telescopes are both limited by technological, design, and
cost constraints related to their optics and detectors. However, for
ground-based observations the FOV is further limited by the area over
which AO correction is possible (see Section~\ref{s:AOtypes}). This
limits the ability to do certain kinds of science, e.g, wide area
surveys at high-resolution.

\vspace{-0.5truecm}

\paragraph{2.3.3. Stability} Another advantage of space observations is 
the long-term stability of the environment, with no gravity, no
seismic disturbances, no weather, and generally smaller thermal
fluctuations. As a result, the PSF and optical geometric distortions
are extremely stable. This makes the space-environment uniquely suited
for science that requires high photometric or geometric
precision. Absolute photometry is possible from space to levels around
$0.01$ mag, and relative differential photometry to levels of order
$0.0001$ mag. The latter is critical for photometric variation studies
such as for planet transits (as in the Kepler mission). The
photometric quality for ground-based AO is limited by temporal and
spatial (field-dependent) PSF variations, which become more
significant for decreasing wavelength and increasing field size. In
the near-IR, current accuracies of $\sim 0.05$ mag (e.g., Davidge
\etal 2003; Vacca \etal 2007) can likely be reduced to near space
quality with the advent of, e.g., MCAO systems (see
Section~\ref{s:AOtypes}). Ground-based AO observations (e.g., for the
Galactic Center) and space-based observations (e.g., for globular
cluster stars) have both produced superb proper motions results, with
the former benefitting from the small diffraction-limited core, and
the latter from the exquisite geometric stability over large time
baselines.\looseness=-2

\vspace{-0.8truecm}

\subsection{Target Accessibility}
\label{ss:access}

\vspace{-0.2truecm}

\paragraph{2.4.1. Time Sampling} One other obvious advantage of
space, and in particular the L2 Sun-Earth Lagrange point, is that
there is no day-night cycle (this does not apply to low-earth orbit
satellites such as HST). From L2 it is possible to continuously follow
time-variable phenomena and to take continuous deep exposures, without
the need to compare or co-add data from different nights during which
observing conditions may have been different.

\vspace{-0.5truecm}

\paragraph{2.4.2. Sky Coverage}

AO observations require bright guide stars to correct for atmospheric
turbulence. With natural guide stars (NGSs), only $\sim 5-50$\% of the
sky is accessible depending on the galactic latitude, for K-band
imaging at moderate Strehl Ratios ($\sim 20$\%; Frogel \etal 2008).
This limitation has now been mostly overcome through the use of Laser
Guide Stars (LGSs). However, even with an LGS system the sky coverage
is not complete, since it still requires an NGS for the tip-tilt
correction (although this NGS can be several magnitudes fainter than
for a pure NGS system). In general, sky coverage drops with increasing
SR and Galactic latitude. For example, NFIRAOS on TMT is expected to
cover $\sim 50\%$ at the Galactic pole (Ellerbroek, priv.~comm.) and
up to 100\% at low Galactic latitude. Projects and proposals exist to
improve sky coverage further using various techniques, both on TMT and
on smaller telescopes (e.g., Keck, CFHT).

\vspace{-0.8truecm}

\section{AO: Current State-of-the-Art and Future Prospects} 
\label{s:AOtypes}

\vspace{-0.2truecm}

The limitations imposed by atmospheric turbulence can be characterized
by the Fried diameter $r_0$, turbulence lifetime $\tau_0$, and
isoplanetic angle $\theta_0$. All are proportional to $\lambda^{6/5}$.
For example, typical magnitudes for a good groundbased site are $r_0 =
10$ cm, $\tau_0 = 6$ ms, and $\theta_0 = 1.8''$ at 5000{\AA} and $r_0
= 160$ cm, $\tau_0 = 95$ ms, and $\theta_0 = 30''$ at $5\mu$m.


For a given SR, the number of subapertures that must be corrected is
proportional to $(D/r_0)^2$ in a time interval $\tau_0$. Consequently
the technology challenges of both applying a sufficiently high
fidelity wavefront correction, and making a sufficiently accurate
wavefront measurement within a turbulent cell of scale length $r_0$ in
a time interval $\tau_0$ are much more manageable for AO corrections
in the near-IR than at optical wavelengths. The number of photons
required for a constant level of correction is roughly $\propto (r_0^2
\tau_0 \lambda)^{-1} = \lambda^{-23/5}$ (the $r_0^2 \tau_0$ comes from
the area of the subaperture and time scale over which a given number
of photons are required; the $\lambda$ comes from the need to achieve
the same path error expressed in waves). Hence, lower wavelengths
require much larger laser power for wavefront sensing.\looseness=-2

Today, AO for near-IR observations is in routine operation on the
Keck, VLT, Gemini and other telescopes. A detailed overview of current
AO capabilities was provided by Frogel \etal (2008, 2009), together
with a roadmap for future development. They found that both AO
performance and the number of AO-enabled refereed science papers has
grown steadily over the last five years. We discuss the various types
of AO in turn below.

\vspace{-0.5truecm}

\paragraph{classical AO} In classical AO, a single guide star, 
either natural (NGS) or laser (LGS), is used to measure the
deformations of the incoming wavefront with a wavefront-sensor and to
correct them with the help of a single Deformable Mirror (DM).  In
this case the FOV is limited to a few times the isoplanetic angle
$\theta_0$. Since $\theta_0 \lta 10''$ for $\lambda < 2 \mu$m,
this is an important limit for many scientific applications. One
limitation of classical AO with an LGS is the so-called called "cone
effect". This results from the incomplete sampling of the turbulence
in front of the aperture due to the finite distance of the LGS. This
becomes increasingly problematic for larger telescope diameters and
shorter wavelengths.

\vspace{-0.5truecm}

\paragraph{LTAO} In Laser Tomography Adaptive Optics (LTAO), multiple 
LGSs and wavefront sensors are used to measure the full volume of
turbulence above the telescope in order to solve for the cone effect
and operate a single conjugate deformable mirror, as in classical AO.
This is what is planned for visible 10m-class AO systems, and for
near-IR first light use on 30m-class ground-based telescopes such as
the TMT. The SR will range from 0.3 at $1.0\mu$m to 0.8 at $2.5\mu$m
over a FoV of about $30"$, with a technical FoV for guide star
acquisition of $2'$. The short wave cutoff for science will be $\sim
0.8 \mu$m, where one might still expect a diffraction limited core at
very low SR (Ellerbroek, priv. comm.).

\vspace{-0.5truecm}

\paragraph{MCAO} An exciting recent advance has been the 
advent of Multi-Conjugate Adaptive Optics (MCAO), in which multiple
DMs are optically conjugated at different altitudes. Multiple
wavefront sensors use LGSs to tomographically measure and compensate
for turbulence-induced phase aberrations in three dimensions. This new
technique increases the compensated FOV, provides a more uniform PSF
over the field, and also solves for the ``cone effect''. The solar
community has been using MCAO to deliver arc-minute scale, fully
corrected near-IR images of the solar granulation. A demonstrator at
the ESO/VLT has delivered stable arc-minute scale images with $\lta
10$\% PSF variations. The first MCAO on Gemini will deliver
nearly-uniform performance over a $2'$ FOV with SR ranging from 45\%
to 80\% from 1--$2.5 \mu$m.

\vspace{-0.5truecm}

\paragraph{GLAO} Ground-Layer Adaptive Optics (GLAO) is a ``lite'' form of 
MCAO, in which only the ground-layer turbulence is corrected. This
technique does not provide the same image quality improvement as MCAO.
It is therefore sometimes referred to as ``seeing enhancement'', with
FWHM improvements of a factor of a few. However, the technique offers
the advantage of a large FOV (up to several arcmin). Moreover, it has
the potential to be implemented on 30m-class ground-based telescopes
at visible wavelengths.

\vspace{-0.5truecm}

\paragraph{MOAO} In Multi-Object Adaptive Optics (MOAO), a number of objects 
are selected in the field. Each object goes through it own AO system
with one DM per object. The correction is based on tomographic
knowledge and is open-loop. Like MCAO, MOAO provides the potential for
significant FOV enhancement over the isoplanetic angle. Its concepts
are well developed and have been run on a testbed, but remain to be
demonstrated on large astronomical telescopes.\looseness=-2

\vspace{-0.5truecm}

\paragraph{Visible-Light AO} Even in the optical, improvements in
technology have demonstrated correction around bright sources across a
narrow (1--$2''$) FOV. These advances have so far been led by
investments by the US Air Force on the 3.5m SOR telescope. In the next
decade we should see these technologies (and others pursued on, e.g.,
CFHT/VASAO and Palomar/PALM3K) move to large ground-based telescopes,
as the cost-effectiveness of high-density DMs and high-power lasers
improve (e.g., Dekany \etal 2009). However, the delivered SRs will
always decrease strongly towards lower wavelengths. Moreover, the
technological challenges to extend the FOV remain enormous. Because
the guide star brightness requirement scales with $\lambda^{-23/5}$,
challenging amounts of laser power will be necessary for visible AO
(independent of $D$). Solutions exist to mitigate these power
requirements, for example using predictive control or uplink
correction (Gavel \etal 2008). Laser power drives the cost of these
systems. Routine visible-light AO operation with {\it both significant
SR and FOV} on 30m-class ground-based telescopes is not foreseen in
the coming decade(s).\looseness=-2

\vspace{-0.5truecm}
  
\paragraph{ExAO} In Extreme AO (ExAO), narrow field, high-SR 
systems are specifically designed to detect planets around nearby
bright stars, as discussed in Section~\ref{s:contrast}.

\vspace{-0.5truecm}

\paragraph{Wavefront Control in Space} Although space has no turbulent 
atmosphere to correct for, figure control is definitely important.
Following in the footsteps of JWST, future observatories are
increasingly likely to use wavefront-sensing for optimum image
quality. Although corrections can be done at much lower speeds that
from the ground, the necessary technologies (e.g., MEMS) have strong
overlap with those (being) developed from ground-based
AO.\looseness=-2

\vspace{-0.8truecm}

\section{Signal-to-Noise Ratio and Limiting Magnitude for Deep/Faint Science}
\label{s:SN}

\vspace{-0.2truecm}

A crucial consideration in the comparison between ground-based and
space-based observations is the atmospheric sky background emission
$B_{\lambda}$ (see Figure~1), which becomes particularly important in
observations designed to reach the faintest possible limits. The noise
contribution in imaging observations is determined by the photometric
aperture size $r$. For seeing-limited observations $r \propto {\rm
FWHM}$, independent of $D$, while for observations with a diffraction
limited core $r \propto \lambda/D$. For given target flux and
wavelength or passband, this yields for the limit in which background
noise dominates other sources of noise that
\begin{equation}
  S/N \propto (D/{\rm FWHM}) \sqrt{\eta/B_{\lambda}} \qquad {\rm or} \qquad
  S/N \propto D^2 \> {\rm SR} \> \sqrt{\eta/B_{\lambda}} ,
\end{equation}
for seeing-limited or diffraction-limited observations,
respectively. Here $\eta$ is the product of the throughput and quantum
efficiency of the system, which nowadays is close to unity for both
ground- and space-based systems. These equations make it clear that
the deepest science benefits are derived from larger telescopes,
better image quality, and lower background.


\begin{figure}[t]
\null
\vspace{-0.7truecm}
\noindent\epsfxsize=0.47\hsize
\epsfbox{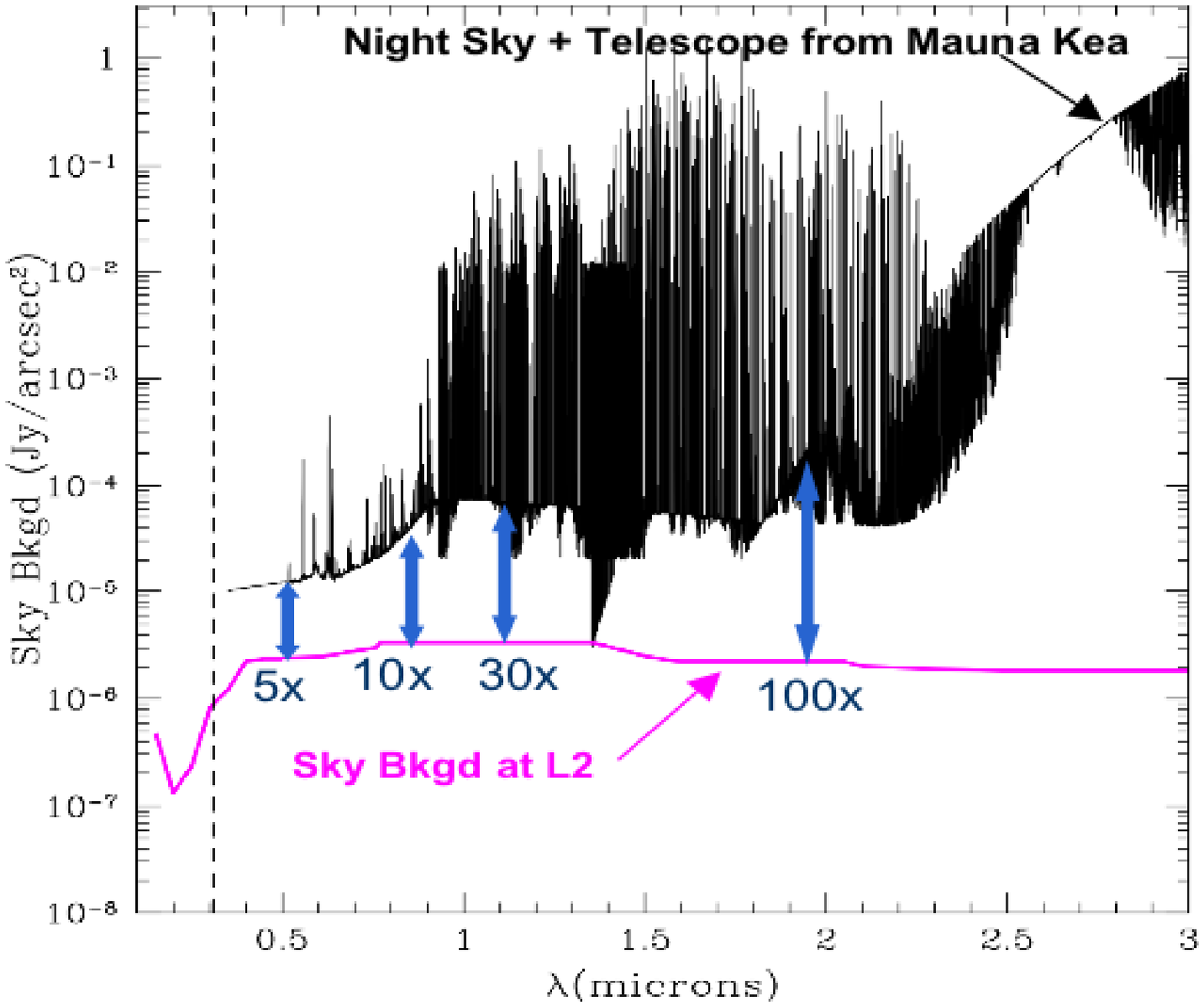}\hfill
\epsfxsize=0.525\hsize
\epsfbox{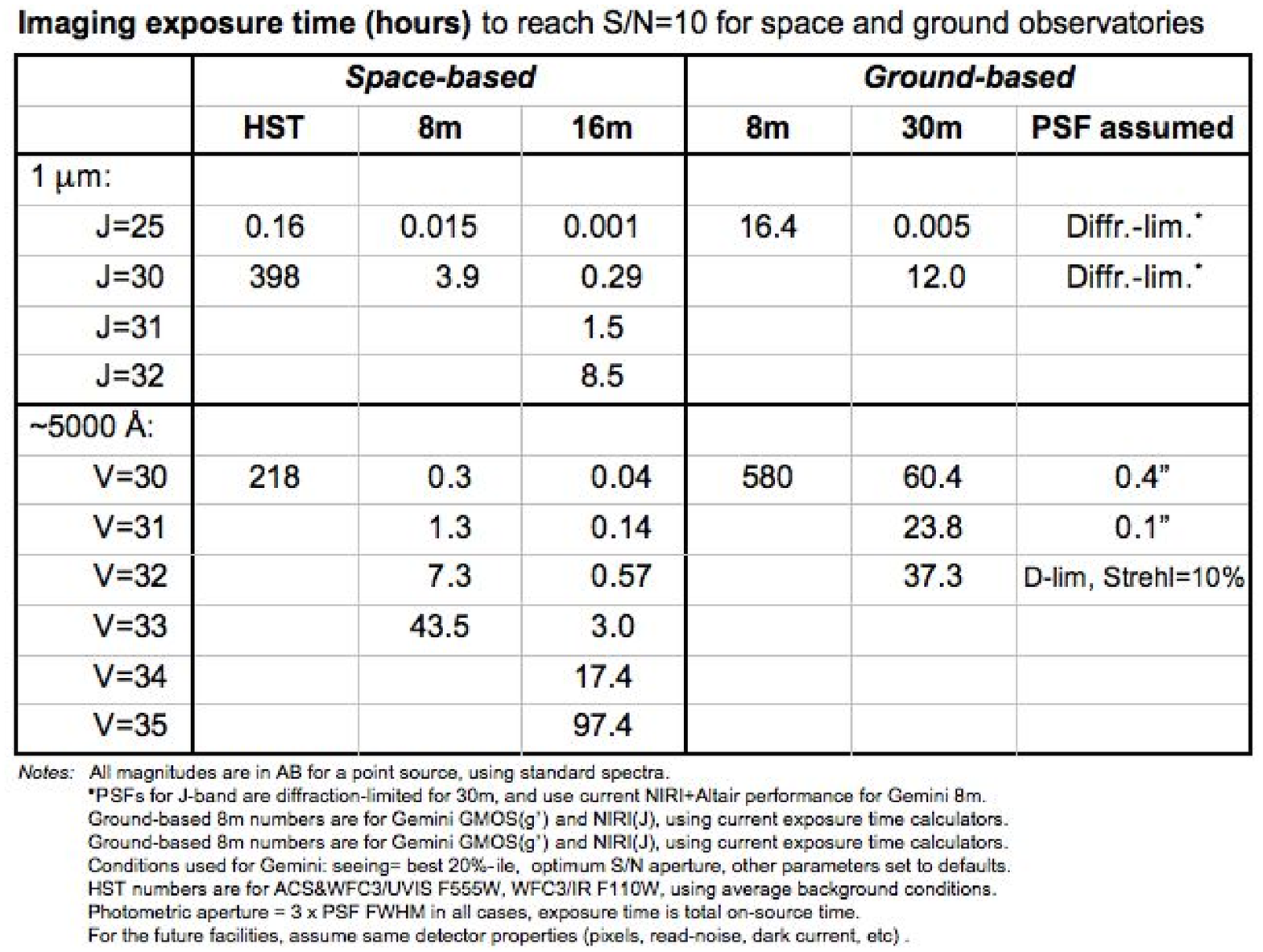}
\null
\vspace{0.5truecm}
\begin{minipage}{\hsize}\footnotesize\baselineskip=4pt 
{\bf Figure~1 (left):} The background flux (from atmospheric
molecular, ionic and continuum emission and telescope thermal
emission) for a ground-based telescope on Mauna Kea as compared to a
space-based telescope at L2.\\ \\ {\bf Table~1 (right):} Sensitivity
comparison between space-based (L2) and ground-based
observatories. The imaging {\it observing time in hours} is listed to
reach $S/N = 10$ at the magnitude listed in the left column. See table
notes and main text for details on the calculations. For the
ground-based 30m, each line uses a different PSF, as listed in the
last column. For the $V$-band, this spans the range of what may be
achievable.\vspace{-0.8truecm}
\end{minipage}
\end{figure}


\vspace{-0.8truecm}

\subsection{Imaging Studies}
\label{ss:SNim}

\vspace{-0.2truecm}

For extremely faint sources (e.g., $V > 30$), the atmospheric sky
background emission, even at the best ground-based sites, is at least
$10^4$--$10^6$ times brighter than the source. To illustrate what is
feasible in this context, Table~1 shows the exposure times needed to
reach a given $S/N$ for a set of representative target magnitudes. The
exposure times are given for the $V$ and $J$-bands at $\lambda =
5000${\AA} and $1\mu$m, respectively. This brackets the range of
wavelengths on which we focus in the present paper. These values were
obtained from the integration or exposure time calculators for 2.4m
(HST), 8m and 16m space-based observatories, the latter two located at
L2, as well as 8m and 30m ground-based observatories with AO
capability. For the ground-based 8m we used existing capabilities, in
particular Gemini/NIRI+Altair in the $J$-band and a Gaussian FWHM of
$0.4''$ (good seeing) in the $V$-band (at present, there are no
general-user optical AO systems available on 8m class telescopes). For
the ground-based 30m telescope in the $J$-band we optimistically
assumed diffraction-limited performance with the same SR as for the
space-based systems. For the $V$-band we present a sampling of three
potential PSFs that span the range of where we might be in a decade or
two: (a) a Gaussian FWHM of $0.4''$ (good seeing); (b) a Gaussian FWHM
of $0.1''$ (optimistic estimate with a successful advent of visible
GLAO); or (c) a diffraction limited core with SR = 10\% (optimistic
estimate with a successful advent of visible LTAO, MCAO, or
MOAO).\looseness=-2

The results for the current generation of facilities show that in the
J-band, HST (with the WFC3 camera) goes several magnitudes deeper than
a ground-based 8m, with only a moderate difference in spatial
resolution. In the V-band, both facilities can reach comparable
depths, due to the smaller difference in background at $\sim
5000${\AA} (see Figure~1). However, HST provides the better spatial
resolution. For future facilities, a 16m space-based observatory at L2
will be capable of reaching $J \sim 32.5$ and $V \sim 34$ in
integrations shorter than 1 day. This is a completely unexplored
parameter space. A 30m ground-based telescope will fall short of these
limits by 2--4 mag. The difference is due to a combination of a higher
sky emission and (in $V$) poorer image-quality on the ground. An 8m
telescope in space at L2 still goes deeper in $V$ than a 30m
ground-based telescope by 1--3 mag. However, in $J$ the depth is more
similar and the ground-based telescope with diffraction-limited AO
will have the better spatial resolution.

Figure~2 provides a graphical way to look at results of calculations
of this nature as function of wavelength, with the spectrum of the
earth as it would be seen at 20pc overplotted.\looseness=-2


\begin{figure*}[t]
\null
\vspace{-1.0truecm}
\noindent\epsfxsize=0.45\hsize
\epsfbox{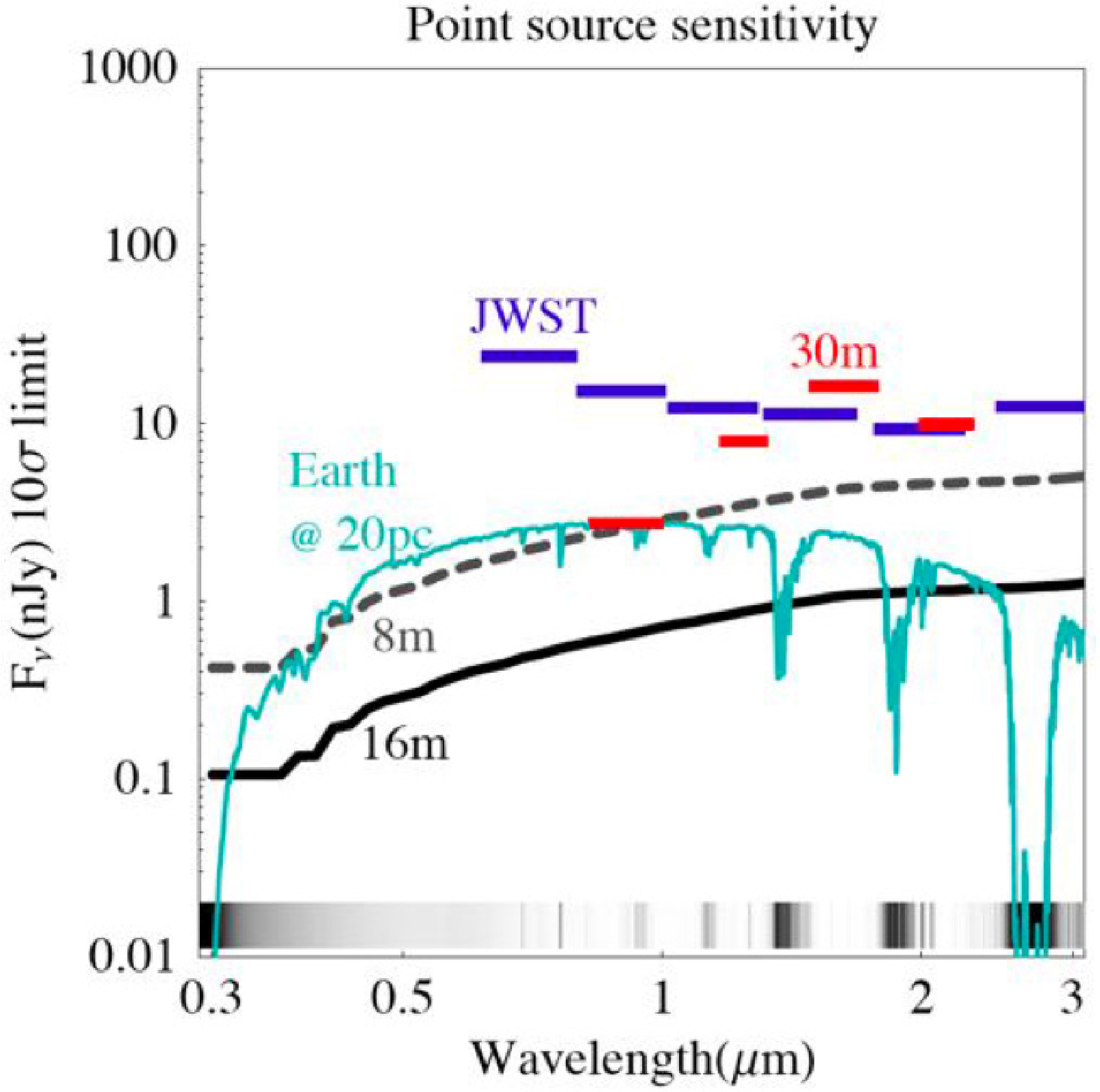}\hfill
\epsfxsize=0.45\hsize
\epsfbox{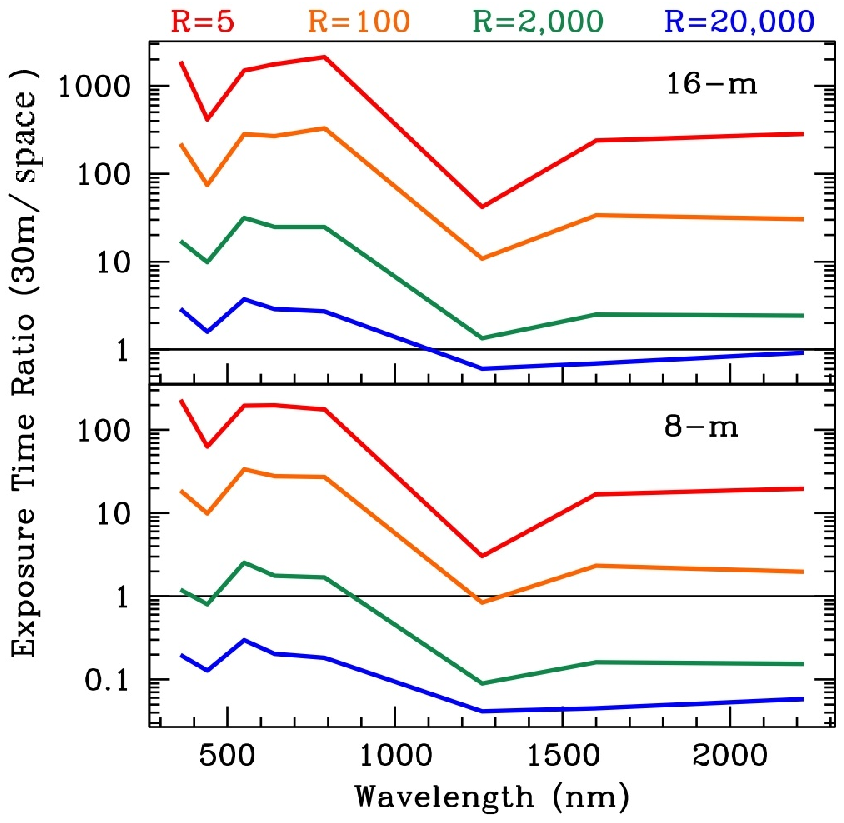}
\begin{minipage}{\hsize}\footnotesize\baselineskip=4pt 
{\bf Figure~2 (left):} 10-sigma point source sensitivities (nJy) for
background-limited 1-hour broadband (R=5) imaging for: an 8-m and 16-m
space telescope (black); a 30-m ground-based telescope with
diffraction-limited AO (red); and JWST (dark blue). Sensitivities in
this figure were calculated with the methodologies of Beckwith
(2008). The superposed spectrum (light blue) shows for comparison a
terrestrial exoplanet at a distance of 20 pc. To detect such a planet
one also has to contend with additional backgrounds not shown here:
(1) the exo-zodi light around the parent star (a factor $\sim 2$
larger than the L2 background itself); and (2) the PSF halo of the
parent star (depends on achieved contrast; see
Section~\ref{s:contrast}). The band at the bottom of the plot shows
the ground-based sky absorption (black = sky opaque to external
radiation).\\ {\bf Figure~3 (right):} Ratio of exposure times for
reaching $S/N = 10$ as function of wavelength $\lambda$, with a
ground-based 30m telescope or an L2 space-based telescope of size 16m
(top) or 8m (bottom), respectively. Ratios are indicated for four
different spectral resolutions ${\cal R}$, with the top curve in each
panel being the lowest resolution.
\end{minipage}
\end{figure*}


\vspace{-0.8truecm}

\subsection{Spectroscopic Studies}
\label{ss:SNspec}

\vspace{-0.2truecm}

Similar considerations apply to spectroscopic observations, since the
spatial resolution determines the smallest slit width and spatial
aperture dimension for producing a final summed spectrum. As an
example, Figure~3 compares the time to reach $S/N=10$ for a point
source as a function of spectral resolution for a 30-m ground-based
telescope, versus the time required for a space-based 8m and 16m
observatory located at L2.  As in Table~1, all are assumed to have the
same instrument and detector performance. In these calculations we
have assumed that the 30m ground-based telescope is
diffraction-limited at wavelengths longer than {\onem}, and is seeing
limited ($0.4''$ FWHM) at lower wavelengths. Some form of optical AO
would decrease the advantage of the space-based observatory, as in
Table~1.\looseness=-2

Figure~3 shows that an 8m space-based telescope is 10 to 100 times
faster than a ground-based 30m for all optical seeing-limited imaging
($R \sim 5$) and up to 40 times faster for most low-resolution ($R
\sim 100$) spectroscopy. The space-based 8m is also more sensitive for
medium-resolution optical spectroscopy ($R \sim 2000$). Similarly, a
16m space telescope is much faster for all spectroscopy in the visible
compared to a seeing-limited 30m ground-based telescope. Wide-field
optical imaging and moderate resolution spectroscopy at the depth
allowed by a new large space-based telescope will allow important new
studies of, e.g., planets (Kasting \etal 2009), distant and/or faint
galaxies (Giavalisco \etal 2009), and resolved stars outside the Local
Group (Brown \etal 2009).

At highest spectral resolutions, background noise ceases to be the
dominant noise contribution, so that space observatories loose their
edge. Also, another important metric is how many sources can be
observed spectroscopically at any given time. Such multiplexing is
possible in space (e.g., JWST/NIRSpec), but ground-based observatories
may well have fewer constraints in pushing this to its limits.

\vspace{-0.8truecm}

\section{High-contrast science}
\label{s:contrast}

\vspace{-0.3truecm}

\subsection{Extreme AO with Coronagraphs}

\vspace{-0.2truecm}

\paragraph{ExAO} Extreme Adaptive Optics pushes high-contrast observations, 
geared in particular towards study of exoplanets. It gives a very high
correction ($SR >90$\%) within a small FOV (a few arcsec,
corresponding to a few AUs to a few tens of AUs around nearby
stars). The target star itself is used for wavefront sensing,
therefore eliminating any anisoplanetic errors. ExAO requires large
numbers of corrected modes, and fast correction rates in order to
achieve the very high image quality. Currently, ExAO systems are
designed and built for the near infrared where the correction is
easier than in the visible (see Section~\ref{s:AOtypes}).

\vspace{-0.5truecm}

\paragraph{Coronagraphs and high-contrast calibration} In order 
to produce high-contrast images, ExAO needs to be combined with
coronagraphs and additional calibration schemes (active calibration,
speckle nulling, differential imaging, etc.). Many coronagraphs have
been developed and proven in the laboratory. The best laboratory
experiments have reached $\sim 5 \times 10^{-10}$ contrast in medium
bands (10\%), which proves the feasibility of the concept for the
detection of Earth analogs. Coronagraphs work best for perfect
non-obstructed apertures.  Although some schemes exist to mitigate the
effects of a central obstruction or segmentation, the performance is
severely affected by the diffraction of these geometrical features,
especially with the typically large central obstruction and wide
support structures required on future 30m-class ground-based
telescopes (hereafter referred to as Extremely Large Telescopes,
ELTs). It is not clear whether internal coronagraphs can be designed
to deliver $10^{-10}$ contrast with large on-axis segmented
telescopes, but the task is facilitated by the increased angular
resolution of ELTs. The coronagraphs need only reach this contrast
level at 10--20 resolution elements ($\lambda/D$) for the detection of
an Earth-twin, as opposed to a few for smaller telescope. After
coronagraphy, space-observations and ground+ExAO-observations require
similar levels of calibration, in order to improve the performance and
remove residual starlight propagation artifacts (speckles). This task
is made easier in space because of the greater thermal and mechanical
stability of a space-based telescope.

\vspace{-0.5truecm}

\paragraph{State of the Art} On the ground, new instruments are being 
built for current 8m-class telescopes and will start operating in
2011. These instruments (e.g., GPI, SPHERE, Subaru, Palomar) involve
ExAO, advanced coronagraphy, spectrographs and polarimeters. These
projects will focus on observations of young ($<2$ Gyr) giant planets
in the near-IR (0.9--$2.5 \mu$m). ELTs also envision high-contrast
instruments, but very little funding has been invested so far. Their
AO is more challenging, and mirror segmentation increases the
difficulty for coronagraphic efficiency and stability. The science
goals for high-contrast ground-based ELTs are very exciting and mostly
include the study of young planets in star forming regions, mature
reflected light giant planets, known radial velocity (RV) planets,
Neptunes for nearby stars, and high-resolution images of
protoplanetary disks.

\vspace{-0.8truecm}

\subsection{The Limits for Ground-Based High-Contrast Science}

\vspace{-0.2truecm}

\paragraph{Science Drivers} Imaging and spectroscopic characterization 
of exoplanets or disks is the principal science motivation for
high-contrast observations. The most exciting goal is arguably the
direct detection of habitable terrestrial planets and the search for
spectroscopic biomarkers. Strong motivations exist for doing these
observations at short wavelengths ($\lta 1\mu$m), as discussed in,
e.g., Kasting \etal (2009) and Lawson \etal (2009). This is because of
the strong oxygen band at 760nm, as well as other potential biomarkers
(ozone, vegetation red edge, Rayleigh scattering). This prompts the
question whether such a science program could be achieved with a
ground-based 30m class telescope. We use the Sun-Earth system at 10 pc
as a template for addressing this discussion.

\vspace{-0.5truecm}

\paragraph{Idealized Analysis} Adaptive optics 
is fundamentally limited by the capacity to analyze the wavefront
because of the finite number of photons available for wavefront
sensing. Based on the analysis by Guyon (2005) this allows one to
derive fundamental limits for high-contrast imaging in the visible for
a 30m class telescope. We consider a perfect telescope with a perfect
AO system and perfect coronagraphic instrumentation. We consider an
ideal wavefront sensor making theoretically optimal use of all
incoming photons (a few existing WFS concepts do offer this level of
sensitivity, but have not yet been deployed on telescopes).  We only
consider errors due to photon noise in the wavefront sensor, and
assume that the AO system has no other source of error (perfect DM, no
calibration error). We assume that WFS is performed with the science
detector, therefore removing any chromatic or non-common path
issues. We also assume correction of both phase aberrations and
amplitude (scintillation) by means of two perfect DMs.  For each
spatial frequency, an optimal exposure time (and therefore optimal
control loop rate) exists which optimizes the time lag effect on the
corrected phase and the photon noise.  Better control, including for
example predictive methods would improve these fundamental limits by a
factor of a few but would not change the main conclusions. With these
ideal assumptions, the residual aberrations from the atmosphere create
a halo in the final raw image with the raw contrast given by
\begin{equation}
   C(\alpha) = 2.348
               { { v^{2/3} \Psi(\alpha/\lambda) \lambda_0^{2/3} } \over
                 { D^2 F^{2/3} r_0^{5/9} \alpha^{5/9} \lambda^{1/9} } } .
\end{equation}
Here $v$ is the wind speed, F the star flux, $\alpha$ the angular
separation, $\Psi$ is a term resulting from Fresnel propagation
between the turbulent layers, and the Fried $r_0$ is defined at
$\lambda_0 = 0.5\mu$m (see Guyon 2005 for derivation and details).

\vspace{-0.5truecm}

\paragraph{Theoretical Limit} We assume a 30m telescope with such 
a perfect system operating at the wavelength of the oxygen A-band
(760nm). The raw final image PSF consists of a residual halo as shown
in Figure 4. This does not correspond to the actual contrast
sensitivity, since this residual halo can be subtracted further using
calibration and differential methods. The ultimate sensitivity limit
is set by the photon noise in this halo. As a template, we consider
the case of the Sun-Earth system at 10pc, with star magnitude m=4.1 (I
band). We also assume 100\% throughput and that a perfect noiseless
halo can be subtracted. The required contrast for our template science
case is $10^{-10}$ at 0.1 arcsec. We find that the exposure time
required to reach $S/N=5$ with a resolution R=70 is $\sim 10^{6}$s
($\gta 10$ days). Given the unrealistic set of assumptions adopted, we
believe that this demonstrates that this observation is not feasible.
 
\vspace{-0.5truecm}

\paragraph{Technological limitations} There are numerous technical 
difficulties that will make it impractical to get close to the
theoretical limit on an ELT in visible light. This includes the
required number of actuators, achievable loop rate, implementation of
science camera wavefront sensing, coronagraph designs for segmented
apertures, stability required for calibration and halo subtraction,
etc. Even at longer wavelengths, the $10^{-10}$ contrast regime will
be extremely challenging. A number of studies of ExAO instruments for
30m class telescopes found that in practice one might reach $10^{-8}$
contrast at $\sim 40$ mas and $10^{-9}$ contrast at 100 mas in the
near-IR ($1.6 \mu$m) (e.g., Macintosh \etal 2006; Cavarroc \etal 2006;
Kasper \etal 2008). Although this comes short of what is required for
studying terrestrial planets, these instruments will have extremely
interesting and exciting scientific capabilities for the broader study
of exo-planetary systems.


\begin{figure}[t]
\null
\vspace{-0.7truecm}
\noindent\epsfxsize=0.4\hsize
\epsfbox{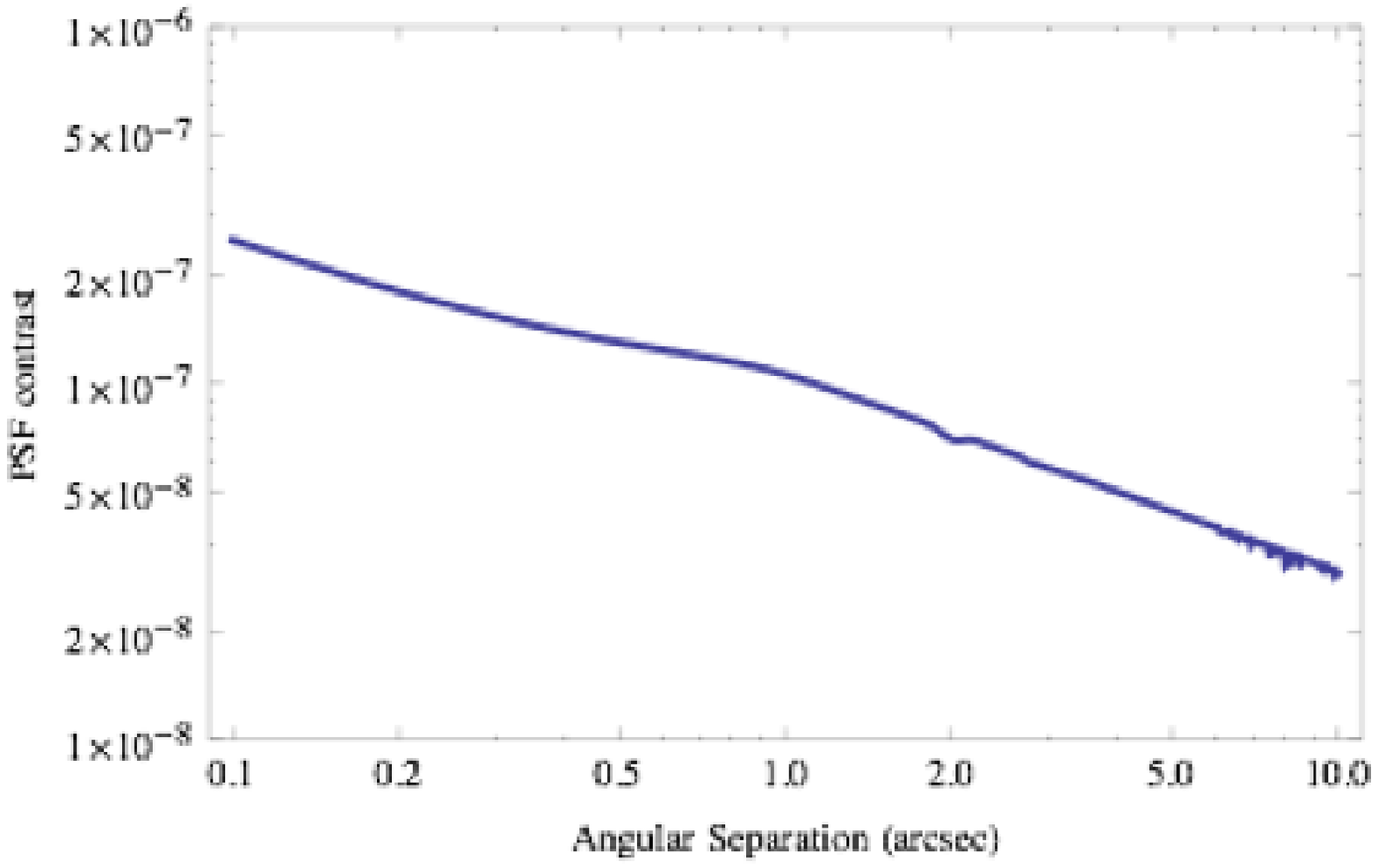}\hfill
\begin{minipage}{0.56\hsize}\vspace{-4.0truecm}\footnotesize\baselineskip=4pt 
\noindent {\bf Figure 4:} Raw PSF for a 30m telescope observing at
760nm with perfect telescope and ExAO. The PSF is shown {\it after}
correction with a perfect coronagraph, and is normalized to be ${\rm
PSF}=1$ at the center {\it without} coronagraph. It therefore shows
the achieved contrast before application of any further differential
methods. Ideal WFS is assumed to be performed with the science
detector at the same wavelength to eliminate non-common path errors or
chromatic propagation effects. The turbulence parameters are
standard. As discussed in the text, even such an idealized setup would
not be sufficient for low-resolution spectroscopy of an earth-twin at
10pc.
\end{minipage}
\end{figure}


\vspace{-0.8truecm}

\subsection{High-Contrast Science from Space}

\vspace{-0.2truecm}

When combining fundamental limits and realistic technical capabilities
for AO on the upcoming 30m class telescopes, it is clear that space
observations will be necessary for very high-contrast science at
wavelengths shorter than {\onem}, as required for the study of
terrestrial planets in the habitable zone of nearby stars. Several
space-based high-contrast projects are under study at this time (e.g.,
Cash \etal 2009a,b; Kasdin \etal 2009; Postman \etal 2009). The most
exciting prospect is to address the habitability of terrestrial
planets and search for life by identifying spectroscopic biomarkers
(O$_2$, O$_3$, H$_2$0, CH$_4$, etc.; Kasting \etal 2009).\looseness=-2

\vspace{-0.8truecm}

\section{Concluding Remarks}
\label{s:conc}

\vspace{-0.2truecm}

Adaptive optics has made exceptional advances in approaching
space-like image quality at higher collecting area, although the exact
prospects for optical-wavelength applications remain uncertain. This
will provide exciting new access to scientific problems that were
previously inaccessible. Nonetheless, there will continue to be
subjects that can {\it only} be studied at optical wavelengths from
space, including wide-field imaging and (medium-resolution)
spectroscopy at the deepest magnitudes, and the exceptional-contrast
observations needed to characterize terrestrial exoplanets and search
for biomarkers. So we expect that there will continue to be a strong
complementarity between observations from the ground and from space in
the coming decades. This provides strong motivating for continued
technology development and new facilities, both for ground- and
space-based optical (and UV) applications.\looseness=-2

\vspace{-0.8truecm}

\section{References}
\label{s:refs}

\smallskip

\frenchspacing\scriptsize\baselineskip=5pt\parskip=0pt

\noindent [ASTRO 2010 abbreviations: 
NOI: Notice of Intent; SW: Science Frontiers White Paper; 
SP: State of the Profession Paper]

\vspace{0.4truecm} 

\hsize=0.48\hsize

\vbox to 4truecm{

\pap Beckwith, S. V. W. 2008, ApJ, 684, 1404

\pap Brown, T., et al. 2009, ``The History of Star Formation in
          Galaxies'', SW

\pap Cash, W., et al. 2009, NOI \#149 (NWO)

\pap Cash, W., et al. 2009, NOI \#120 (JWST Starshade)

\pap Cavarroc, C., et al. 2006, in Modeling, Systems Engineering, and
          Project Management for Astronomy II. Cullum \etal eds., Proc.~SPIE, 
          6271, 18

\pap Davidge, 2003, PASP, 115, 635

\pap Dekany, R., et al. 2009, NOI \#40 

\pap Frogel, J. A., et al. 2008, ``AURA's Assessment of Adaptive
          Optics: Present State and Future Prospects (v4.0)''
          (http://www.aura-astronomy.org/nv/Astro2010Panel\\Docs/AURAs\
          assessment\ of\ AO\ V4.pdf)

\pap Frogel, J. A., et al. 2009, ``Frontier Science and Adaptive
          Optics On Existing and Next Generation Telescopes'', SP


\pap Gavel, D., et al. 2008,  in Adaptive Optics Systems. Hubin
          \etal eds., Proc.~SPIE, 7015, 8

\pap Gehrels, N., et al. 2009, NOI \#73 (JDEM)

\pap Giavalisco, M., et al. 2009, ``The Quest for a Physical
          Understanding of Galaxies Across the Cosmic Time'', SW

}

\vspace{-4.1truecm}
\hsize=2.0833\hsize

\vbox to 4truecm{

\papp Guyon, O. 2005, ApJ, 629 (revised: astro-ph/0505086)


\papp Kasdin, N.J., et al. 2009, NOI \#132 (THEIA)

\papp Kasper, M., et al. 2008, in Adaptive Optics Systems. Hubin
          \etal eds., Proc.~SPIE, 7015, 46

\papp Kasting, J. 2009, ``Exoplanet Characterization and the
          Search for Life'', SW

\papp Lawson, P. R., et al. 2009, ``Exoplanet Community Report''
          (http://exep.jpl.nasa.gov/exep\_exfCommunityReport.cfm)

\papp Levine, M., et al. 2009, NOI \#133 (TPF-C)

\papp Macintosh, B., et al. 2006, in Advances in Adaptive Optics
          II, Ellerbroek, \etal eds., Proc.~SPIE, Vol.~6272, 20

\papp McCarthy, P. J., et al. 2009, NOI \#143 (GMT)

\papp Postman, M., et al. 2009, NOI \#13 (ATLAST)


\papp Stone, E. C., et al. 2009, NOI \#158 (TMT)

\papp Vacca, W. D., Sheehy, C. D., \& Graham, J. R. 2007, ApJ,
          662, 272

\papp Verinaud, C. et al., 2006, in Advances in Adaptive Optics II, 
          Ellerbroek, \etal eds., Proc.~SPIE, Vol.~6272, 19

}


\end{document}